\documentclass[twocolumn,pre]{revtex4}

\usepackage{dcolumn}
\usepackage{amsmath}

\usepackage{graphicx}

\newcommand{\etal}{{\it{}et~al.}}

\begin{document}

\title{Prediction of highly cited papers}
\author{M. E. J. Newman}
\affiliation{Department of Physics and Center for the Study of Complex
  Systems, University of Michigan, Ann Arbor, MI 48109}

\begin{abstract}
  In an article written five years ago, we described a method for
  predicting which scientific papers will be highly cited in the future,
  even if they are currently not highly cited.  Applying the method to real
  citation data we made predictions about papers we believed would end up
  being well cited.  Here we revisit those predictions, five years on, to
  see how well we did.  Among the over 2000 papers in our original data
  set, we examine the fifty that, by the measures of our previous study,
  were predicted to do best and we find that they have indeed received
  substantially more citations in the intervening years than other papers,
  even after controlling for the number of prior citations.  On average
  these top fifty papers have received 23 times as many citations in the
  last five years as the average paper in the data set as a whole, and 15
  times as many as the average paper in a randomly drawn control group that
  started out with the same number of citations.  Applying our prediction
  technique to current data, we also make new predictions of papers that we
  believe will be well cited in the next few years.
\end{abstract}

\maketitle

\section{Introduction}
Citations of scientific papers are considered to be an indicator of papers'
importance and relevance, and a simple count of the number of citations a
paper receives is often used as a gauge of its impact.  However, it is also
widely believed that citations are affected by factors besides pure
scientific content, including the journal a paper appears in, author name
recognition, and social effects~\cite{Cronin84,Leydesdorff98}.  One
substantial and well-documented effect is the so-called cumulative
advantage or preferential attachment bias, under which papers that have
received many citations in the past are expected to receive more in future,
independent of content.  A simple mathematical model of this effect was
proposed by Price~\cite{Price76}, building on earlier work by
Yule~\cite{Yule25} and Simon~\cite{Simon55}, in which paper content is
ignored completely and citation is determined solely by preferential
attachment plus stochastic effects.  Within this model, the expected number
of citations a paper receives is a function only of its date of
publication, measured from the start of the topic or body of literature in
which the paper falls, and shows a strong ``first-mover effect'' under
which the first-published papers are expected to receive many more
citations on average than those that come after them.  Indeed the variation
in citation number as a function of publication date is normally far wider
than the stochastic variation among papers published at the same time.  In
a previous paper~\cite{Newman09a} we compared the predictions of Price's
model against citation data for papers from several fields and found good
agreement in some, though not all, cases.  This suggests that pure citation
numbers may not be a good indicator of paper impact, since much of their
variation can be predicted from publication dates, without reference to
paper content.

Instead, therefore, we proposed an alternative measure of impact.  We
proposed that rather than looking for papers with high total citation
counts, we should look for papers with counts higher than expected given
their date of publication.  Since publication date is measured from the
start of a field or topic, and since different topics have different start
dates, one should only use this method to compare papers within topics.
The appropriate calculation is to count the citations a paper has received
and compare that figure to the counts for other papers on the same topic
that were published around the same time.  In our work we used a simple
$z$-score to perform the comparison: we calculate the mean number of
citations and its standard deviation for papers published in a window close
to the date of a paper of interest, then calculate the number of standard
deviations by which that paper's citation count differs from the mean.
Papers with high $z$-scores we conjecture to be of particular interest
within the field.

One promising feature of this approach is that the papers it highlights are
not necessarily those with the largest numbers of citations.  The most
highly cited papers are almost always the earliest in a field, in part
because of the first-mover effect but also because they have had longer to
accumulate citations.  More recent papers usually have fewer citations, but
they may still have high $z$-scores if the count of their citations
significantly exceeds the average among their peers.  Thus the method allows
us to identify papers that may not yet have received much attention but
will do so (we conjecture) in the future.

In our previous study, we used this method to identify some specific papers
that we believed would later turn out to have high impact.  Here we revisit
those predictions to see whether the papers identified have indeed been
successful.  To quantify success, we look again at citation counts, and to
minimize the preferential attachment bias we compare them against randomly
drawn control groups of papers that had the same numbers of citations at
the time of the original study.  As we show, our predictions appear to be
borne out: the papers previously identified as potential future successes
have received substantially more attention than their peers over the
intervening years.

\section{Previous results}
In this section we briefly review some of the results from our previous
paper~\cite{Newman09a}, which we will refer to as Paper~1.

In Paper~1 we examined citation data from several different fields in
physics and biology, but made specific predictions for one field in
particular, the field of interdisciplinary physics known as network
science.  This field is an attractive one for study because it has a clear
beginning---a clear date before which there was essentially no published
work on the topic within the physics literature (although there was plenty
of work in other fields)---and a clear beginning is crucial for the theory
developed in the paper to be correct and applicable.  (It is also the
present author's primary field of research, which was another reason for
choosing it.)  It is again on papers within network science that we focus
here.

In Paper~1 we assembled a data set of 2407 papers on network science
published over a ten year period, starting with the recognized first
publications in 1998 and continuing until 2008.  The data set consisted of
papers in physics and related areas that cited one or more of five
highly-cited early works in the
field~\cite{WS98,BA99b,AB02,DM02,Newman03d}, but excluding review articles
and book chapters (for which citation patterns are distinctly different
from those of research articles).  We then calculated the mean and standard
deviation of the number of citations received by those papers as a function
of their time of publication within a moving window.  Crucially, however,
our theoretical studies indicate that ``time'' in this context is most
correctly defined not in terms of real time, but in terms of number of
papers published within the field.  If $n$ papers have been published in
the field in total (with $n=2407$ in this case), then the ``time''~$t$ of
publication of the $i$th paper is defined to be $t=i/n$, where papers are
numbered in the order of their publication.  Thus $t$ runs from a lowest
value of $1/n$ for the first paper in the field to $1$ for the most recent
paper.  It is in terms of this variable that we perform our averages.

\begin{figure}
\begin{center}
\includegraphics[width=\columnwidth]{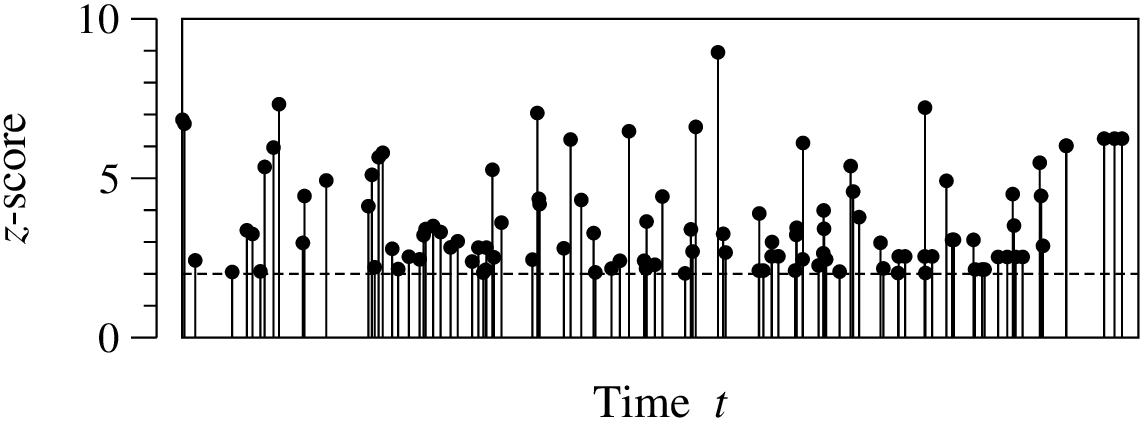}
\end{center}
\caption{A plot of $z$-scores for citations to papers in the data set of
  Ref.~\cite{Newman09a}.  Each $z$-score is equal to the number of standard
  deviations by which the citations to the corresponding paper exceed the
  mean for papers published around the same time.  Only $z$-scores greater
  than two (the dashed line) are plotted.  As described in the text, the
  ``time'' in the horizontal axis is measured by the number of papers
  published, not real time.}
\label{fig:outliers}
\end{figure}

Armed with these results, we then calculate a $z$-score for each paper as
described in the introduction: we take the count of citations received by a
paper, subtract the mean for papers published around the same time, and
divide by the standard deviation.  Figure~\ref{fig:outliers} reproduces a
plot from Paper~1 of $z$-scores for papers in the data set.  Only
$z$-scores that exceed~2 are shown, since these are the ones we are most
interested in, corresponding to papers whose citation counts are
significantly above the mean for their peers.  As the figure shows, there
are within this set a small number of papers that stand out as having
particularly high $z$-scores.  Our suggestion was that these were papers to
watch, that even if they did not currently have a large number of
citations, they would in future attract significant attention.
Table~\ref{tab:winners} lists twenty of the top papers by this measure from
our 2008 data set along with their citation counts and
$z$-scores~\cite{note}.

\begin{table*}
\begin{tabular}{lllrrl}
Year & Author(s) & Reference & 2008 cites & 2013 cites & $z$-score \\
\hline
1998 & Watts, D. J. and Strogatz, S. H. &	Nature \textbf{393}, 440             &	2623 &	7807 & 6.834 \\
1999 & Barab\'asi, A.-L. and Albert, R. &	Science \textbf{286}, 509            &	2579 &	7955 & 6.713 \\
2002 & Milo, R. \etal                   &	Science \textbf{298}, 824            &	620 &	1775 & 5.964 \\
2002 & Ravasz, E. \etal                 &       Science \textbf{297}, 1551           &  563 &	1343 & 5.358 \\
2004 & Barrat, A. \etal                 &	Proc. Natl. Acad. Sci. \textbf{101}, 3747 &	233 &	760 & 5.798 \\
2005 & Guimer\`a, R. and Amaral, L. A. N. &     Nature \textbf{433}, 895             &  138 &	677 & 5.266 \\
2005 & Palla, G. \etal                  &	Nature \textbf{435}, 814             &	137 &	902 & 7.047 \\
2005 & Santos, F. C. and Pacheco, J. M. &	Phys. Rev. Lett. \textbf{95}, 098104 &	74 &	458 & 6.216 \\
2006 & Newman, M. E. J.                 &	Proc. Natl. Acad. Sci. \textbf{103}, 8577 &	65 &	901 & 8.954 \\
2006 & Ohtsuki, H. \etal                &	Nature \textbf{441}, 502             &	63 &	437 & 6.614 \\
2006 & Zhou, C. S. \etal                &	Phys. Rev. Lett. \textbf{96}, 034101 &	59 &	310 & 6.475 \\
2006 & Kim, P. M. \etal                 &	Science \textbf{314}, 1938           &	35 &	222 & 6.109 \\
2007 & Dosenbach, N. U. F. \etal        &	Proc. Natl. Acad. Sci. \textbf{104}, 11073 &	15 &	371 & 7.214 \\
2007 & Palla, G. \etal                  &	Nature \textbf{435}, 814             &	15 &	281 & 5.386 \\
2007 & Tadi\'c, B. \etal                  &	Int. J. Bifurc. Chaos \textbf{17}, 2363 &	6 &	58 & 5.488 \\
2008 & Perc, M. and Szolnoki, A.        &	Phys. Rev. E \textbf{77}, 011904     &	4 &	129 & 6.019 \\
2008 & Kozma, B. and Barrat, A.         &	Phys. Rev. E \textbf{77}, 016102     &	4 &	69 & 6.019 \\
2008 & Hidalgo, C. A. and Rodriguez-Sickert, C. &	Physica A \textbf{387}, 3017 &	1 &	29 & 6.245 \\
2008 & Clauset, A. \etal                &	Nature \textbf{453}, 98              &	1 &	255 & 6.245 \\
2008 & Estrada, E. and Hatano, N.       &	Phys. Rev. E \textbf{77}, 036111      &	1 &	47 & 6.245
\end{tabular}
\caption{Twenty of the papers with the highest $z$-scores in 2008,
  as calculated using the method described in the text, listed in
  chronological order along with their citation counts both now and at the
  time of the previous study~\cite{note}.}
\label{tab:winners}
\end{table*}

Five years have elapsed since Paper~1 was written, giving us five years of
additional citation data for the same papers.  In the following section we
look more closely at the papers in Table~\ref{tab:winners} and other top
papers and show that, while there is plenty of variation in the fortunes of
individual papers, these papers have clearly done better on average than
their peers in the intervening time.

\section{New results}
\begin{figure}[b]
\begin{center}
\includegraphics[width=\columnwidth]{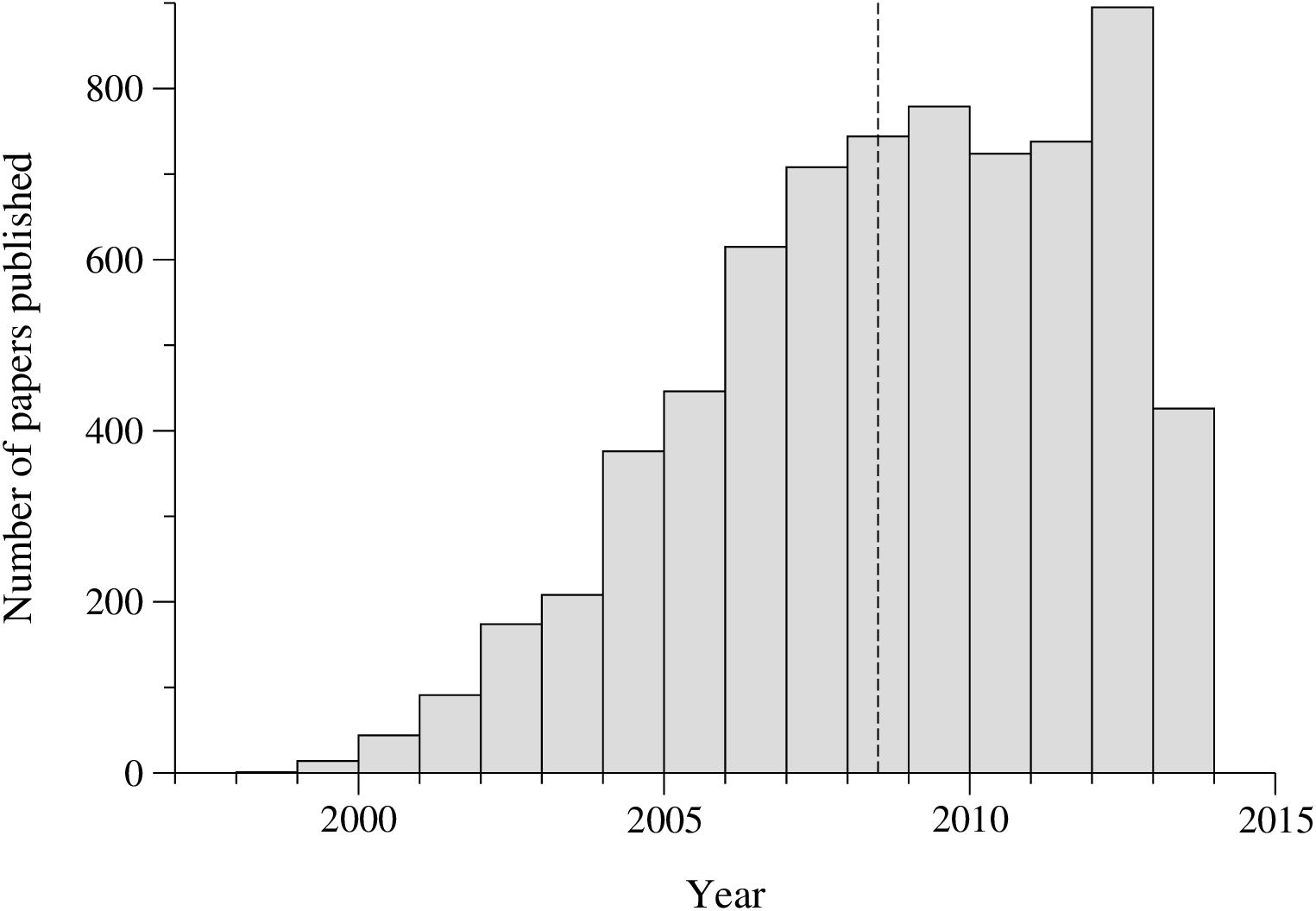}
\end{center}
\caption{The number of papers in our 2013 data set that were published in
  each year.  The dashed line indicates the date of our previous
  study~\cite{Newman09a}.  More than half of the papers represented here
  were published after this date.}
\label{fig:years}
\end{figure}

To quantify the success (or lack of success) of the papers identified by
high $z$-scores in our 2008 study, we once more turn to the citation
record.  Duplicating our earlier methodology, we have again assembled a set
of papers citing the same five early works in the field but excluding
reviews and book chapters.  Where the data of Paper~1 covered only a ten
year period, however, our new data cover an additional five years for a
total of fifteen years from 1998 to 2013.  The new data set contains 6976
papers in total, over twice as many as we found in 2008.  Indeed the first
and most obvious conclusion from the new data is that the field of network
science has grown tremendously in the last five years.
Figure~\ref{fig:years} shows a plot of the number of papers in the data set
by year of publication, and the rapid growth is immediately apparent.  In
the first year of the data set, the year 1998, there was only one paper.
In 1999 there were 14.  In 2012, the last full year represented, there were
895~papers.  Despite this growth however, and despite the fact that 2012
produced a bumper crop of papers, one might tentatively say from the figure
that the field has plateaued---the vigorous growth of the first decade
appears roughly to have leveled off around 800 papers per year after about
2008.

Using our new data set we have located each of the 2407 papers studied in
Paper~1 and determined how many citations, in total, they had received as
of August 2013, approximately five years after the study of Paper~1.  For
the selection of papers in Table~\ref{tab:winners} these figures are given
in the second-to-last column of the table.  From them we can calculate the
gain in citations received by each paper.  As the table shows there is
plenty of variation, but all the papers, without exception, have
significantly more citations than they did five years earlier, and all of
them are now well cited by the standards of the field.  To give some points
of reference, within the entire new data set a citation count of 40 or more
puts a paper in the top 10\%, while a citation count of 295 or more puts it
in the top~1\%.  Thus all but one of the twenty papers in
Table~\ref{tab:winners} fall in the top~10\% and twelve of them fall in the
top~1\%.

Looking more closely at the table we notice that there are some papers that
had very few citations in 2008, some of them only a single citation,
particularly the most recent papers at the bottom of the list.  The
inclusion of these papers in the table is somewhat dubious, since their
large $z$-score rests entirely on the fact of their having received a few
citations very shortly after publication (at which time receiving even a
single citation is a statistically surprising event).  To avoid possible
biases, therefore, we henceforth exclude from our reckoning those papers
that had received less than five citations at the time of our earlier
study.  Most of the results reported in the remainder of this article are
averages for the fifty papers that have the highest $z$-scores after this
culling is performed.

The median number of additional citations received in the last five years,
by all 2407 papers from our earlier study, was~10.  Among our fifty
top-scoring papers the corresponding number is~238.  (We quote median
figures because the distributions are long-tailed and in such cases means
can be strongly dependent on a small number of highly cited papers in the
tail.  The median does not suffer from this problem.)  Thus the papers we
identified have fared very much better than the average.  These results
suggest that our analysis is capable of identifying papers that will on
average receive large numbers of citations in future.

\begin{figure}
\begin{center}
\includegraphics[width=8cm]{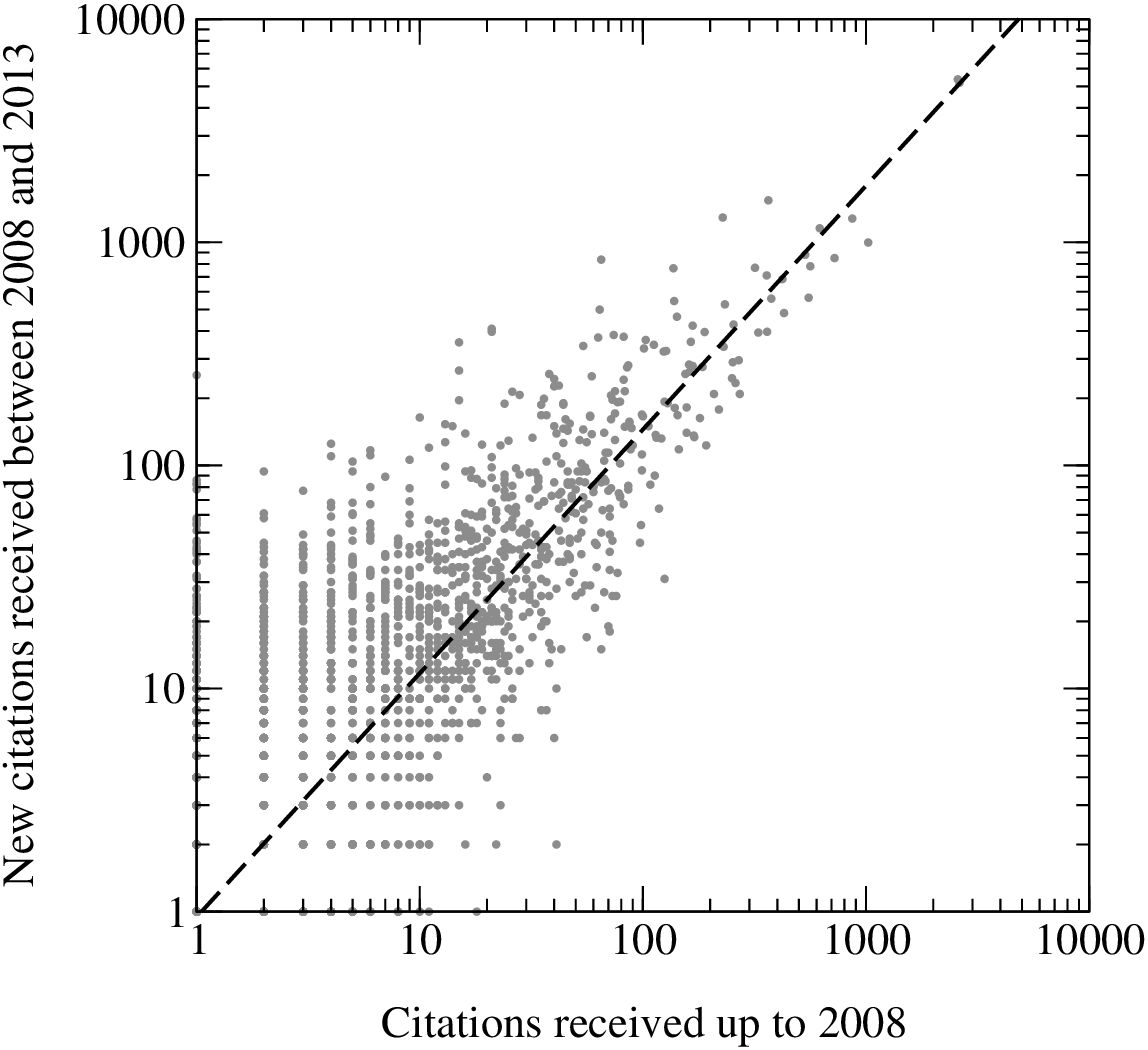}
\end{center}
\caption{Scatter plot of the number of new citations received by papers
  between 2008 and 2013 as a function of the number they had in 2008.
  Logarithmic axes are used to allow for the wide variation in citation
  counts.  The dashed line shows the best least-squares fit to a power law
  (which appears as a straight line on logarithmic axes).}
\label{fig:prefattach}
\end{figure}

This on its own, however, is not as impressive a feat as it may sound.  As
discussed in the introduction, the citation process is thought to display
preferential attachment, meaning that papers that have received many
citations in the past are expected to receive many in the future as well,
and there is evidence in support of this hypothesis in our data.
Figure~\ref{fig:prefattach} shows a scatter plot of the number of new
citations received by the papers in our study in the last five years
against the number they had at the time of Paper~1 in 2008.  As the plot
shows, there is a strong positive correlation between the two numbers.  If
we merely wanted to predict papers that would receive many citations,
therefore, it would be easy to do---we just pick papers that have large
current numbers of citations.  Indeed if we look at the fifty papers from
our data set that had the largest numbers of citations back in 2008, we
find that the median number of new citations they received in the last five
years is~376, even more than the figure of 238 for the fifty papers
identified by our $z$-score analysis.

But this misses the point.  The point of our prediction method was to
filter out preferential attachment, to identify promising papers even if
they had not yet received many citations.  For instance, in Paper~1 we
singled out the 2006 article by Ohtsuki~\etal~\cite{OHLN06} as a paper that
we believed would receive substantial attention in future, even though at
that time it had received only 63 citations, a prediction that seems to
have been borne out, given that the paper has received an additional 374
citations in the five years since.

One crude way to determine how successful we have been at picking out
promising papers of this kind is to calculate not the difference in
citation numbers between our first and second measurements, but the ratio.
Since the preferential attachment model predicts that the number of new
citations received will be proportional to the old number, our null
hypothesis under the model is that the ratio of new to old citations should
be the same, apart from stochastic fluctuations, for all papers.  Thus we
can tentatively identify papers with higher-than-average ratios as having
outperformed our expectations.

Looking again at all papers and calculating this new-to-old ratio for each
one, we find a median value of~2.00---the median paper had twice as many
citations in 2013 as it did in 2008.  For the fifty promising papers picked
out by our analysis the corresponding figure is more than twice as large,
at~5.24.  Conversely, the figure for the fifty overall highest-cited papers
is just 2.48---only a little greater than the figure for all papers in the
data set.  In other words if our goal is to pick papers that will
outperform the preferential-attachment null-model expectation, then the
strategy of picking papers that already have many citations is a poor
one---on average these papers barely beat a random basket of papers.  The
papers identified by our analysis, on the other hand, do significantly
better than average, increasing their citation scores by a factor much
greater than that of the typical paper.

This analysis is still not entirely satisfactory, however, since it relies
on the assumption of linear preferential attachment.  Previous studies
suggest that attachment may be nonlinear in practice~\cite{Newman01d,JNB03}
and this appears to be true of our data---the best power-law fit to the
data in Fig.~\ref{fig:prefattach} (shown as the dashed line in the figure)
has an exponent somewhat greater than~1.  And there's no reason to assume a
power law to be the best functional form.  It is possible that the
attachment probability could vary in quite a complex manner with the number
of citations, the attachment law being substantially different for
rarely-cited papers from what it is for well-cited ones.

A more robust way to test the citation performance of the papers identified
in our analysis is to compare them against a control group, or many control
groups, consisting of papers with similar prior citation performance.  This
allows us to answer the question, ``Do the papers we have identified
perform better than other papers with the same number of citations?''  To
this end, we have taken the complete collection of 2407 articles studied in
Paper~1 and from them sampled 100 random subsets of papers using Markov
chain Monte Carlo.  Each of these sets contains fifty papers, like the
original set identified in our $z$-score analysis, and each has the same
total number of citations as the original set, which is~$10\,356$.  Apart
from this constraint, however, the sets are drawn at random.

\begin{table*}
\begin{tabular}{lllrl}
Year & Author(s) & Reference & 2013 cites & $z$-score \\
\hline
1998 & Watts, D. J. and Strogatz, S. H. & Nature \textbf{393}, 440 & 7807 & 7.043 \\
1999 & Barab\'asi, A.-L. and Albert, R. & Science \textbf{286}, 509 & 7955 & 7.183 \\
2002 & Milo, R. \etal &	Science \textbf{298}, 824 & 1775 & 6.893 \\
2004 & Newman, M. E. J. and Girvan, M. & Phys. Rev. E \textbf{69}, 026113 & 1519 & 9.871 \\
2005 & Palla, G. \etal & Nature \textbf{435}, 814 &  902 & 9.074 \\
2005 & Guimer\`a, R. and Amaral, L. A. N. & Nature \textbf{433}, 895 &  683 & 9.142 \\
2005 & Santos, F. C. and Pacheco, J. M. & Phys. Rev. Lett. \textbf{95}, 098104 &  458 & 7.030 \\
2006 & Newman, M. E. J. & Phys. Rev. E \textbf{74}, 036104 &  431 & 9.763 \\
2007 & Fortunato, S. and Barth\'elemy, M. & Proc. Natl. Acad. Sci. \textbf{104}, 36 &  418 & 9.148 \\
2007 & Dosenbach, N. U. F. \etal & Proc. Natl. Acad. Sci.  \textbf{104}, 11073 &  371 & 7.312 \\
2007 & Palla, G. \etal & Nature \textbf{446}, 664 &  281 & 8.441 \\
2008 & Blondel, V. D. \etal & J. Stat. Mech. \textbf{2008}, P10008 &  287 & 8.924 \\
2008 & Rosvall, M. and Bergstrom, C. T. & Proc. Natl. Acad. Sci. \textbf{105}, 1118 &  258 & 7.806 \\
2008 & Clauset, A. \etal & Nature \textbf{453}, 98 &  255 & 7.365 \\
2008 & Santos, F. C. \etal & Nature \textbf{454}, 213 &  226 & 7.769 \\
2010 & Buldyrev, S. V. \etal & Nature \textbf{464}, 1025 &  192 & 9.002 \\
2010 & Good, B. H. \etal & Phys. Rev. E \textbf{81}, 046106 &  119 & 8.270 \\
2010 & Centola, D. & Science \textbf{329}, 1194 &  114 & 7.935 \\
2011 & Liu, Y. Y. \etal & Nature \textbf{473}, 167 &  126 & 9.289 \\
2012 & Gao, J. X. \etal & Nature Physics \textbf{8}, 40 & 32 & 7.130
\end{tabular}
\caption{Twenty of the papers on network science with the highest $z$-scores
  in 2013.  We predict these papers will receive substantially more
  citations over the next few years than other papers from the same field
  with similar current numbers of citations~\cite{note}.}
\label{tab:newwinners}
\end{table*}

Looking at the papers in these 100 sets, we now find the number of new
citations each received in the last five years, the difference between
their citation counts in 2013 and 2008.  Calculating the median new
citations for each set, we find the average figure over all sets to be
$15.7\pm0.4$, a much lower number---by a factor of~15---than the median of
238 measured for the fifty leaders from our analysis.  Alternatively, as
before, we can compute the ratio of new to old citations, again taking a
median for each set, and we find an average figure of $2.23\pm0.02$, not
very different from the $2.00$ we found for the entire data set (perhaps
suggesting that our assumption of linear preferential attachment was a
reasonable one).  As reported above, the equivalent figure for the set of
fifty leading papers was more than twice the size, at~5.24.

By these measures it appears that the predictions of Paper~1 are quite
successful.  Papers identified using our method outperformed by a wide
margin both the field at large and randomly drawn control groups that
started out with the same number of citations.

\section{New predictions}
Encouraged by these results and capitalizing on the fact that we have a new
data set of papers and their citations up to the year 2013, we now apply
our methods to the new data to make predictions about papers that will be
highly cited in the next few years.

Working with the entirety of our new 6976-paper data set, we again
calculate the mean and standard deviation of the numbers of citations
received as a function of time and look for the papers with the highest
$z$-scores---those that exceed the mean for their publication date by the
largest number of standard deviations.  Table~\ref{tab:newwinners} is the
equivalent of Table~\ref{tab:winners} for this new analysis, listing twenty
of the papers with the highest $z$-scores within the field of network
science.

A few observations are worth making.  First, note that the numbers of
citations received by these papers are substantially greater than those
received by the papers in Table~\ref{tab:winners} at the time of our first
round of predictions.  All but one of them have over 100 citations already,
putting them in the top few percent of the data set.  This is probably in
part a sign of the rapid growth of the field mentioned earlier.  A more
rapid rate of publication means more citations are being made, and hence
more received, particularly by the most prominent papers.  It's worth
emphasizing, however, that each of the papers in the table earned its place
by receiving significantly more citations, by many standard deviations,
than other papers in the same field published around the same time.  So the
citation counts are not merely high, but anomalously so.  Moreover, the
$z$-scores in Table~\ref{tab:newwinners} are significantly higher than
those in Table~\ref{tab:winners}, so the margin by which the top papers
outperform expectations has also grown.  Second, we note that while some of
the papers in the list would be unsurprising to those familiar with the
field of network science, such as the influential early papers by Watts and
Strogatz~\cite{WS98} and by Barab\'asi and Albert~\cite{BA99b}, there are
also some more recent papers listed, such as the papers by
Buldyrev~\etal~\cite{Buldyrev10} on interdependent networks or
Liu~\etal~\cite{LSB11} on controllability, which received two of the
highest $z$-scores in the table.  Our analysis suggests that these papers
will have an outsize impact a few years down the road, relative to what one
would expect given only their current numbers of citations.  Third, we
notice the abundance in the list of papers published in the journal
\textit{Nature}, and to a lesser extent \textit{Science} and
\textit{Proceedings of the National Academy of Sciences}.  The publication
venues of the papers in our old list, Table~\ref{tab:winners}, were more
diverse.  The predominance of these journals could just be
coincidence---the sample size of 20 is small enough to make such a thing
plausible.  But it is also possible that it is a real effect.  Perhaps
these journals have over the last few years established a special name for
themselves as attractive venues for publication of work in this particular
field.  Or perhaps there has been a deliberate change in editorial policy
that has resulted more papers in the field being accepted for publication
in these journals.  Alternatively, these journals may have pulled ahead of
their competition in the accuracy of their peer review, so that they are
better able to identify (and hence publish) papers that will be important
in the field.  Or it may be that papers in these high-profile journals tend
to be cited more than papers in other journals because they are more
visible, and hence they are more likely to appear in our table.  Or the
truth might be a combination of all of these factors, and perhaps some
others as well.

\section{Conclusions}
In this paper we have revisited predictions made in 2008 of scientific
papers in the field of network science that, according to those
predictions, should receive above average numbers of citations, even though
they may not yet have had many citations at that time.  Looking at those
predictions five years on, we find that the papers in question have indeed
received many more citations than the average paper in the field.  Indeed
they have received substantially more even than comparable control groups
of randomly selected papers that were equally well cited in 2008.  Because
of the so-called preferential attachment effect, one can quite easily
identify papers that will do well just by looking for ones that have done
well in the past.  But the papers identified by our analysis did well even
when one controls for this effect, indicating that the analysis is capable
of identifying papers that will be successful not merely because they
already have many citations.  We hope, though we cannot prove, that the
additional factors driving this success are factors connected with paper
quality, such as originality and importance of research and quality of
presentation.

We have also applied our methods to contemporary data, from 2013, to make a
new round of predictions, identifying papers that, according to the metrics
we have developed, are expected to see above average citation in the next
few years, when compared to other papers that currently have the same
number of citations.  It will be interesting to see whether these
predictions in turn come true.

\begin{acknowledgments}
  This work was funded in part by the US National Science Foundation under
  grants DMS--0405348 and DMS--1107796.
\end{acknowledgments}

\end{document}